\documentclass[aps,pra,twocolumn,superscriptaddress]{revtex4-1}
\usepackage{graphicx}
\usepackage{amsmath,amssymb}

\newcommand{\vett}[1]{\mathbf{#1}}

\renewcommand{\i}{\imath}

\begin{document}

\title{Nonlinear Gamow vectors, shock waves and irreversibility in optically nonlocal media}

\author{Silvia Gentilini} 
\affiliation{Institute for Complex Systems, Via dei Taurini 19, 00185 Rome (IT)}

\author{Maria Chiara Braidotti} 
\affiliation{Department of Physics, University Sapienza, Piazzale Aldo Moro 5, 00185 Rome (IT)}
\affiliation{Institute for Complex Systems, Via dei Taurini 19, 00185 Rome (IT)}

\author{Giulia Marcucci} 
\affiliation{Department of Physics, University Sapienza, Piazzale Aldo Moro 5, 00185 Rome (IT)}

\author{Eugenio DelRe} 
\affiliation{Department of Physics, University Sapienza, Piazzale Aldo Moro 5, 00185 Rome (IT)}

\author{Claudio Conti} 
\affiliation{Department of Physics, University Sapienza, Piazzale Aldo Moro 5, 00185 Rome (IT)}
\affiliation{Institute for Complex Systems, Via dei Taurini 19, 00185 Rome (IT)}
\date{\today}

\begin{abstract}
  Dispersive shock waves dominate wave-breaking phenomena in Hamiltonian systems. In the absence of loss, these highly irregular and disordered waves are potentially reversible.
However, no experimental evidence has been given about the possibility of inverting the dynamics of a dispersive shock wave and turn it into a regular wave-front.
Nevertheless, the opposite scenario, i.e., a smooth wave generating turbulent dynamics is well studied and observed in experiments.

Here we introduce a new theoretical formulation for the dynamics in a highly nonlocal and defocusing medium described by the nonlinear Schroedinger
equation. Our theory unveils a mechanism that enhances the degree of irreversibility.
This mechanism explains why a dispersive shock cannot be reversed in evolution even for
an arbitrarirly small amount of loss.

Our theory is based on the concept of nonlinear Gamow vectors, i.e., power dependent generalizations of the counter-intuitive and hereto elusive exponentially decaying states in Hamiltonian systems. 
We theoretically show that nonlinear Gamow vectors play a fundamental role in nonlinear Schroedinger models: they may be used as a generalized basis for describing the dynamics of the shock waves, and affect the degree of irreversibility of wave-breaking phenomena. Gamow vectors allow to analytically calculate the amount of breaking of time-reversal with a quantitative agreement with numerical solutions.

We also show that a nonlocal nonlinear optical medium may act as a simulator for the experimental investigation of quantum irreversible models, as the reversed harmonic oscillator.

\end{abstract}
\maketitle

\section{Introduction}
One of the fundamental problems in physics is the description of irreversible processes.
Irreversible processes are denoted by exponentially decaying observables associated with a preferential arrow of time.
However, exponential dynamics of the wavefunction can be excluded from first principles in leading theoretical models, such as the Hilbert space (HS) formulation of quantum mechanics (QM).
In this formulation, there are theorems stating that the solution of the Schr\"odinger equation cannot have an exponential evolution. \cite{Khalfin57, Hegerfeldt94}

To explain damping, one introduces loss in a phenomenogical way, accounting for the coupling with the environment.\cite{abohm99,Civitarese04}
However, exponential time decay of observable quantities, 
which signals irreversible dynamics, 
is routinely detected in experiments, even when the role of the environment is negligible.

This is known as the {\it probability problem} and is common to quantum and classical systems.
\cite{Prigogine93}

A paradigmatic example of the issues related to irreversibility in classical systems is found in nonlinear wave propagation. In regimes allowing for shock waves, singular solutions may be excited by smooth initial conditions that generate an evolution dominated by non-regular, highly disordered, waves (wave-breaking). \cite{Gentilini2013}
The evolution of smooth initial conditions towards a highly disordered and incoherent regime 
is intuitive. However, the reverse process, albeit possible, is strongly puzzling, and against well accepted ideas about entropy.

Dispersive shock waves (DSW) and wave-breaking in Hamiltonian models therefore question reversibility. They seem to indicate a preferential 
arrow of time that should be excluded from first principles, because no coupling with the environment is foreseen. \cite{Hoefer2006,Trillo11,Garnier13,Gentilini:14,ContiGrelu14,Moro14,FleisherPicozziCondensation}

Here we pose the following question: is there a connection between shock waves and the problem of the irreversibility of the wavefunction?
These two problems are seemingly unrelated, but if one carefully looks at the dynamics of a shock wave (as detailed below), one realizes that exponential
dynamics can be identified. We show in the following that the specific case of a nonlocal nonlinearity enables to unveil the link and lay the foundations for further developments.

The starting point of the analysis is that one can circumvent the absence of exponentially decaying solution for the wavefunction by enlarging the Hilbert space for including non-normalizable wavefunctions that decay exponentially in evolution.
This is known as the rigged Hilbert space (RHS) formulation of quantum \cite{ABohm81} and classical mechanics  \cite{Castagnino97, Chruscinki02}.
Exponential dynamics are explained in terms of the excitation of states with complex energy eigenvalues.
These states, known as Gamow vectors (GVs), were originally introduced in nuclear physics, and model irreversibility in Hamiltonian systems. \cite{Gamow28, Sudarshan78, Parravicini80, ABohm98}
Even if GV are not-normalizable, they can be used as a generalized basis,
as it happens, e.g., for plane waves.
However, the watermark of GV is a quantized decay rate. This quantization is radically different from the standard quantization of real eigenvalues, and involve the imaginary part of the GV eigenvalues.
Notably enough,  so far, no experimental evidence of the quantization of decay rates has been reported. We also remark that this has nothing to do with quantum many-body theory, or out-of-equilibrium systems (see, e.g., \cite{Civitarese04,degrootbook}).

In this manuscript, we show that GV can be introduced to describe highly nonlinear regimes in nonlocal media.  We show that the solution of the relevant nonlinear Schroedinger equation in the presence of shock waves can be given in terms of GV with power-dependent, quantized decay rates.
GV are shown to quantitatively describe shock waves, in terms of the amplitude and the phase of the complex wavefunction and they explain the observed exponential dynamics.

We also demonstrate that the excitation of GV radically changes the degree of irreversibility of the dynamics.
Specifically, the theoretical tools arising from the use of GV enable to explicitily consider and calculate the degree of irreversibility,
defined as the difference between the forward and the backward propagated wavefunction. This calculation is found to be
in quantitative agreement with numerical solutions, and shows that even an arbitrary small amount of loss produces a breaking of time-reversal enhanced by shock waves. 

This manuscript is organized as follows: in section \ref{sec:nonloc} we introduce the leading model;
in section \ref{sec:GVexp} we introduce GVs and the resulting expansion;
the specific case of the reversed harmonic oscillator is analyzed in section \ref{sec:RO};
section \ref{sec:nonlocalsolitons} describes the link between the ground state GV, shock waves and nonlocal solitons;
section \ref{sec:decays} describes the exponentially decaying dynamics;
section \ref{sec:irrev} deals with the definition of the degree of irreversibility and the way it can be calculated by the GV expansion,
quantitative comparison with numerical simulations is also reported;
conclusions are drawn in  \ref{sec:conc}.

\section{Nonlocal nonlinearity}\label{sec:nonloc}
We start from the paraxial wave equation for the propagation of an optical beam 
with amplitude $A$ and wavelength $\lambda$ in a medium with refractive index $n_0$,
and linear loss length $L_{loss}$ 
[$\mathbf{R}=(X,Y)$] \cite{Gentilini:14}
\begin{equation}
2\imath k \frac{\partial A}{\partial Z}+\nabla^2_{XY} A+2k^2 \frac{\Delta n[|A|^2](\vett{R})}{n_0} A=-\imath\frac{ k }{ L_{loss}}A\text{.}
\label{paraxial}
\end{equation}
$A$ is normalized such that $I=|A|^2$ is the intensity, $P_{MKS}=\int I d\mathbf{R}$ is the power,
$k=2\pi n_0/\lambda$ is the wavenumber.
In (\ref{paraxial}) $\Delta n$ is the nonlinear nonlocal perturbation to the refractive index, 
\begin{equation}
\Delta n[|A|^2](\vett{R})=n_2 \int G_2(\vett{R}-\vett{R}')I(\vett{R'})d\vett{R}'
\label{nonlocal1}
\end{equation}
with $G_2$ the kernel function, normalized such that $\int G_2 d\mathbf{R}=1$.
The local Kerr effect corresponds to 
 $G_2=\delta(\vett{R}-\vett{R'})$. 
For an exponential nonlocality $G_2(X,Y)=G(X) G(Y)$, 
being $G(X)=\exp(-|X|/L_{nloc})/(2L_{nloc})$, with $L_{nloc}$ the nonlocality length.
\begin{figure}
\includegraphics[width=0.45\textwidth]{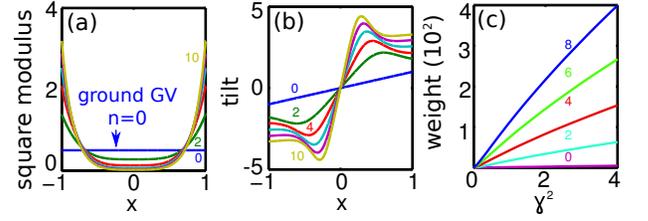}
\caption{
(Color online) (a) GV $|\mathfrak{f}_n^{-}(x)|^2$ for increasing even order $n$; (b)  $\partial_x \arg[\mathfrak{f}_n^{-}(x)]$ for
increasing even order;
(c) weight of the GV expansion of a Gaussian wavepacket.
\label{figuregamow1}}
\end{figure}
\begin{figure}
\includegraphics[width=0.45\textwidth]{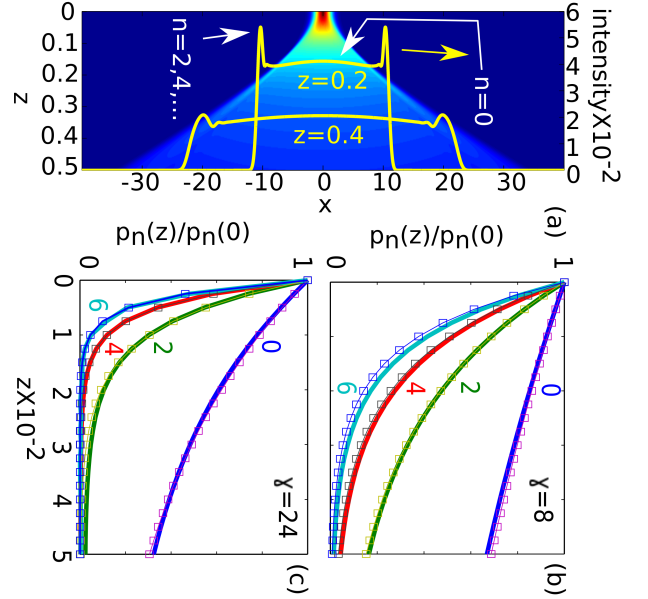}
\caption{
(a) Numerical solution of Eq.(\ref{paraxialnorm}) with $P=10^4$ and $\sigma^2=10$;
(b) projection on GV for increasing $n$ for $\alpha=0.3$ and $\gamma=8$,
continuous lines are from Eq.(\ref{paraxialnorm}), dots are from Eq.(\ref{expprob});
(c) as in (b) for $\gamma=24$.
 \label{coefficientgamow}}
\end{figure}

Since the problem is separable in the highly nonlocal approximation (HNA) considered below,
we consider the case of one transverse dimension $X$.
For a defocusing nonlinearity ($n_2<0$), we write Eq.(\ref{paraxial}) in terms of dimensionless variables, letting $W_0$ be the beam waist, 
$x=X/W_0$, and $z=Z/Z_d$, with $Z_d=k W_0^2$ the diffraction length:
\begin{equation}
  \imath \frac{\partial \psi}{\partial z}+\frac{1}{2}\frac{\partial^2 \psi}{\partial x^2}-P K(x)*|\psi(x)|^2  \psi=-\imath\frac{\alpha}{2}\psi\text{,}
\label{paraxialnorm}
\end{equation}
where $\alpha=Z_d/L_{loss}$, $\psi=A W_0/\sqrt{P_{MKS}}$, and $\langle \psi| \psi\rangle=1$, brackets denoting the HS scalar product.
The asterisk in (\ref{paraxialnorm}) is a convolution, and $P=P_{MKS}/P_{REF}$ with $P_{REF}=\lambda^2/4\pi^2 n_0 |n_2|$, and $K(x)=W_0 G(x W_0)=\exp(-|x|/\sigma)/2\sigma$ with $\sigma=L_{nloc}/W_0$.

\section{Gamow vector expansion}\label{sec:GVexp}
Starting from Eq.(\ref{paraxialnorm}), in the HNA, one obtains an effectively linear Hamiltonian system:
we have \cite{Snyder1997,Folli12}
\begin{equation}
  K*|\psi(x)|^2\cong \kappa(x),
\end{equation}
and
\begin{equation}
  \imath \psi_z=\hat{H}\psi,
  \end{equation} with the Hamiltonian
\begin{equation}\hat{H}=\tfrac{1}{2}\hat{p}^2+V(x),\end{equation} being $V(x)=P \kappa(x)$ and $\hat{p}=-\imath\partial_x$. 

\noindent GVs are numerable generalized eigenvectors of $\hat{H}$ with complex eigenvalues
$\phi_n^G(x)$ with $n=0,1,2,...$; they exist outside 
the HS of Lebesgue integrable functions. In the RHS,
they furnish a generalized basis for a normalizable wavepacket, being $\hat{H}\phi_n^G=E_n \phi_n^G$,
\begin{equation}
\psi(x,0)=\sum_{n=0}^{\infty} \phi_n^G(x)+\phi_b(x)=\phi^G(x)+\phi_b(x)\text{.}
\label{exp1}
\end{equation}
In (\ref{exp1}) $\psi(x,0)$ is the initial state, the discrete part is {\it not} the usual spectrum of normalizable bound states with real valued eigenvalues (which is not present here), but an ensemble of vectors with complex eigenvalues
\begin{equation}
  E_n=E_n^R-i\Gamma_n/2\text{,}
\end{equation}
with real part $E_n^R$, and immaginary part $\Gamma_n$ corresponding to the decay rates. \cite{Sudarshan78,ABohm81,delaMadrid02}
\\$\phi_b$ is the {\it background component}, that is due to the part
of the continuum spectrum of $\hat{H}$ not containing GVs.
The key difference between $\phi^G$ and $\phi_b$ lies in the fact that, upon evolution, the former decays exponentially, while exponential decay is not allowed for $\phi_b$. 
We have
\begin{equation}
\begin{array}{l}
\psi^{G}(x,z)=e^{-i\hat H z}\phi^G(x)=\sum_n \phi_n^G(x) e^{-i E^R_n z-\tfrac{\Gamma_n}{2} z}\text{.}
\end{array}
\label{exp2}
\end{equation}
During evolution GV cannot emerge from the background $\phi_b$,
because they form a semigroup under the action of the evolution operator $e^{-i\hat H z}$ for $z>0$. \cite{ABohm98}

\section{Reversed harmonic oscillator}\label{sec:RO}
In the HNA, for $\alpha=0$, GV can be explicitly written: letting
\begin{equation}
  \kappa(x)=\kappa_0^2-\frac{\kappa_2^2 }{2} x^2,
\end{equation}
we have,
\begin{equation}
  \hat{H}=P\kappa_0^2 +\hat{H}_{ro},
\end{equation}
with $E_n^{R}=P\kappa_0^2$, and
\begin{equation}
  \hat{H}_{ro}=\frac{\hat{p}^2}{2}-\frac{\gamma^2 x^2}{2},
\end{equation}
with $\gamma^2=P\kappa_2^2$. 
For the exponential nonlocality, we have $\kappa_0^2=1/2 \sigma$ and $\kappa_2^2=1/\sqrt{\pi}\sigma^2$.  

\noindent We let $\psi=\exp\left(-\imath \kappa_0^2 P z\right)\phi$, with
\begin{equation}\imath\phi_z=\hat{H}_{ro}\phi\text{.}
  \end{equation}
$\hat{H}_{ro}$ is the Hamiltonian of the reversed harmonic oscillator (R-HO). \cite{Kemble35, Castagnino97,Shimbori99}
In the classical limit, the trajectories of the RHO obey $\dot x(t)=-\gamma x(t)$, i.e.,
they corresponds to a dissipative system, and were also found in the hydrodynamical approximation of DSW. \cite{FolliNJP2013} 
This shows that the nonlocal optical propagation in a defocusing medium is a physical realization of a quantum dissipative system.\cite{Chruscinski2003} GVs for the RHO are
\begin{equation}
\hat{H}_{ro}\mathfrak{f}_n^\pm=\left(\frac{\hat{p}^2}{2}-\frac{\gamma^2 x^2}{2}\right)\mathfrak{f}_n^\pm=E_n^{\pm} \mathfrak{f}_n
\end{equation}
with purely immaginary $E_n^{\pm}=\pm \imath (n+\tfrac{1}{2})$ with \cite{Chruscinski2004}
\begin{equation}
  \mathfrak{f}_n^\pm(x)=\frac{\sqrt[4]{\pm \i \gamma}}{\sqrt{2^n n! \sqrt{\pi}}}  H_n(\sqrt{\pm \i\gamma}x)\exp(\mp \i\frac{\gamma}{2} x^2),
\end{equation}
being $H_n(x)$ the Hermite polynomials.
For any $n$, two GV exist: $\mathfrak{f}_n^{-}$ displaying an exponential decay,  
and $\mathfrak{f}_n^ {+}=(\mathfrak{f}_n^{-})^*$ with exponential growth,
and
\begin{equation}
  \langle \mathfrak{f}_m^{+}|  \mathfrak{f}_n^{-} \rangle=\delta_{mn}\text{,}
\end{equation}
being $\delta_{mn}$ the Kronecker symbol.
$\mathfrak{f}_n^ {+}$ is the time-reversed correspondent of $\mathfrak{f}_n^{-}$.

In Fig.\ref{figuregamow1}a we show the profiles of the lowest order even GV; their modulus squared 
increases with $x$. We have
\begin{equation}
  \hat{H}_{RO}=\sum_{n=0}^\infty E_n^{-} |\mathfrak{f}_n^-\rangle\langle\mathfrak{f}_n^+|
  \end{equation},
and correspondingly
\begin{equation}
\phi^G(x)=\sum_{n=0}^\infty \mathfrak{f}_n^-(x) \langle\mathfrak{f}_n^+|\psi(x,0)\rangle\text{.}
\label{GamowComponents}
\end{equation}
In the RHS theory the initial state $\psi(x,0)$ is prepared before the interaction, 
the state in the future is expressed by the $\mathfrak{f}_n^-$ that decays upon evolution.

\section{Nonlocal solitons and dispersive shock waves}\label{sec:nonlocalsolitons}
In the HNA, the fundamental soliton corresponds to the Gaussian shaped ground state of the HO.\cite{Snyder1997}
The analytical prolongation to imaginary-valued energies of the HO are the GV.
There is hence a precise link between intrinsically irreversible states (the GV) and the 
propagation invariant self-trapped beam (the fundamental soliton):
the prolongation of the ground state of the standard HO (the fundamental soliton) to immaginary eigevalues, furnishes
the fundamental GV $\mathfrak{f}_0^{-}(x)$.

There is also another notable connection with DSW. Not only GVs are found in the same
regime of the nonlocal DSW;\cite{Ghofraniha07} but the phases of the GVs have the signatures that are commonly ascribed to DSW.
In Fig.\ref{figuregamow1}b we show the GV phase profile when increasing $n$, which progressively
resemble DSW (e.g., compare with Fig.1 of \cite{Ghofraniha07}). 
In these terms the singularity that arises upon the generation of shock waves, when increasing the power,
can be interpreted as the excitation of higher order GVs.

\section{Decay processes}\label{sec:decays}
In the probabilistic interpretation of RHS-QM, the projection Eq.(\ref{GamowComponents})
over $\sqrt{\Gamma_n} \mathfrak{f}_n^+$ gives the probability $p_n(z)$ of finding the system in a decaying GV.
This gives the evolution of the irreversible part of the wavefunction, which can be also interpreted as the decaying process of elementary excitations with energy given by $E_n^{R}=\kappa_0^2 P$ and quantized decay rates $\Gamma_n=\gamma(2n+1)$:
\begin{equation}
\psi^G(x,z)=\sum_n^N \langle \mathfrak{f}_n^+|\psi(x,0)\rangle \mathfrak{f}_n^-(x) 
e^{-i \kappa_0^2 P z} e^{-\frac{\Gamma_n}{2} z}
\label{gamowevolution}
\end{equation}
The expansion in Eq.(\ref{gamowevolution}) is only meaningful for $z>0$ and 
\begin{equation}
p_n(z)= \Gamma_n |\langle \mathfrak{f}_n^+|\psi(x,0)\rangle|^2 e^{-\Gamma_n z}
\label{expprob}
\end{equation}
gives the $z-$dependent weight of the $n$-order GV. If $\psi(x,0)$ is a pure GV,\begin{equation}  
  \psi(x,0)=\mathfrak{f}_n^{-}(x),
\end{equation} we have
\begin{equation}
  p_n(z)=\Gamma_n \exp(-\Gamma_n z),
  \end{equation}
with normalization\cite{Shimbori99}
\begin{equation}\int_0^\infty p_n(z)dz=1\text{.}
  \end{equation}
The excited GVs depend on the initial profile. For a Gaussian beam,
\begin{equation}\psi(x,0)=\psi_F(x)=\exp(-x^2/2)/\sqrt[4]{\pi},\end{equation} all the odd terms in Eq.(\ref{gamowevolution}) vanish due the $x-$parity, and for the first two even GV we have
\begin{equation}p_0^G(z)= 2 \frac{\gamma^{3/2}}{\sqrt{1+\gamma^2}} \exp(- \gamma z),\end{equation} and \begin{equation}p_2^G(z)=5 \frac{\gamma^{3/2}}
  {\sqrt{1+\gamma^2}} \exp(-5 \gamma z). \end{equation}
In Fig.~\ref{figuregamow1}c we show the weights $p_n(0)$ for various GV versus the normalized power $\gamma^2=\kappa_2^2 P$: they tend to zero for $P\rightarrow 0$ and grow with $P$.
Their quantized decay rates scale with power as $\gamma\propto\sqrt{P}$.

\noindent We underline that GV are not decay processes depending on the coupling with the environment (i.e., {\it extrinsic}), but exponentially decaying states arising from the time-reversible Hamiltonian ({\it intrinsic} origin).

\noindent To show the occurrence of the GVs in the original nonlinear model,
we solve Eq.(\ref{paraxialnorm}) with $\psi(x,0)=\psi_F(x)$.
At low power (not reported) no exponential decay is found.
At high power (high wave amplitude), in correspondence of the DSW, 
the resulting dynamics clearly display exponential decays (Fig. \ref{coefficientgamow}a). The shape of the beam strongly resembles the excitation of the 
ground state GV, corresponding to a central plateau; and lateral tails can be identified with higher order GVs (compare Fig.~\ref{coefficientgamow}a with Fig.~\ref{figuregamow1}a).
As the GVs decay exponentially and the power is conserved, the beam displays a self-similar evolution, with an exponential spreading following the classical trajectories of the dissipative system.
To provide quantitative evidence of $\mathfrak{f}_n^{-}$ ,
we project the wavefunction $\psi(x,z)$ over $\mathfrak{f}_n^{+}$ and retrieve 
$p_n(z)$ in Eq.(\ref{expprob}), as shown in Fig.\ref{coefficientgamow}b,c.
$p_n(z)$ decays with quantized rates $\Gamma_n=(2n+1)\gamma$.
This trend has been verified for various $\gamma^2$ and initial conditions, and confirms the theoretical analysis. The evidence
of the quantization of decay rates is the most direct signature of the GVs, as shown in Fig. \ref{coefficientgamow}b,c.
Deviations from the exact exponentials are due to the finite degree of nonlocality in Eq.(\ref{paraxialnorm}). 

\section{Gamow vectors and the degree of irreversibility}\label{sec:irrev} 
DSW are formally reversible, but 
in real world experiments there is always an amount of loss. 
The interplay between linear losses and the excitation of the GV
is subtle.
We quantify in the following the way irreversibility, in the presence of linear losses,
is enhanced by the excitation of nonlinear GV.
In order to give a quantitative analysis, we define below the degree of irreversibility as the
scalar between the forward and backward evolution in way such that when this product is zero the dynamics is totally irreversibile. We show below that the degree of irreversibility is enhanced by the nonlinearity with respect to linear loss, and this can quantitative calculated by retaining into account the excitation of GV, and using their exponential dynamics in the analysis.

\begin{figure}
\includegraphics[width=0.45\textwidth]{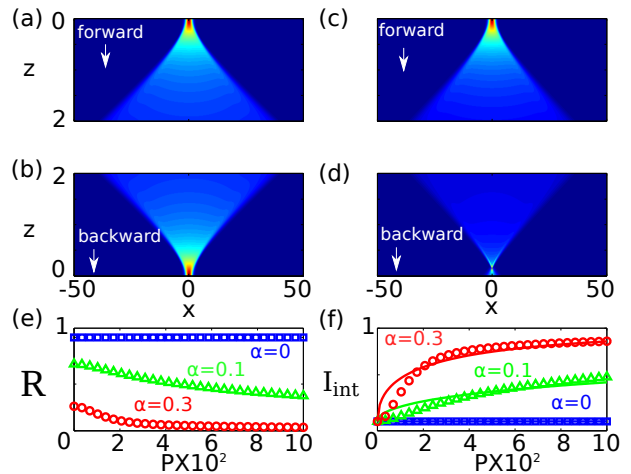}
\caption{
(a) forward progagation of $\psi_F(x)$ for $\alpha=0$; 
(b) backward propagation corresponding to panel (a);
(c) as in (a) for $\alpha=0.3$; 
(d) backward propagation corresponding to panel (b); 
(e) degree of reversibility versus $P$ for various $\alpha$; 
(f) fractional degree of intrinsic irreversibily;
the dashed line is the estimated after Eq.(\ref{irreq}).
Parameters: $L=2$, $\sigma^2=10$
\label{figureirreversible1}}
\end{figure}

\noindent We start showing in Fig.~\ref{figureirreversible1}a-d the numerical solutions of Eq.(\ref{paraxialnorm}). We let $\psi(x,0)=\psi_F(x)$ and retrieve $\psi(x,L)$. We then use the conjugated propagated field as new initial condition $\psi'(x,0)=\psi(x,L)^*$.
As long as dynamics are reversible the propagated $\psi_B(x)=\psi'(x,L)$ coincides 
with $\psi_F(x)$ (Fig. \ref{figureirreversible1}a,b).
Repeating the procedure for $\alpha=0.3$, reversibility does not occur
(Fig. \ref{figureirreversible1}c,d).
The discrepancy between $\psi_F$ and $\psi_B$ increases with $P$, for $\alpha>0$. We introduce the degree of reversibility by
\begin{equation}
  R=|\langle \psi_F|  \psi_B\rangle|^2.
\end{equation}

For a fixed $L$, $R=R(\alpha,P)$, and in the absence of loss $\alpha=0$, the dynamics is exactly reversible
\begin{equation}R(0,P)=1,\end{equation} as show in the figure \ref{figureirreversible1}a,b.
For $\alpha>0$, in the absence of nonlinear effects $P=0$,
the degree of reversibility is given by the linear loss
\begin{equation}R(\alpha,0)=\exp(-\alpha L)\text{,}\end{equation}
increasing linear losses reduces reversibility.
We calculate $R(\alpha,P)$ and, as shown Fig.\ref{figureirreversible1}e, the breaking of reversible character appears more pronounced when increasing $P$.

\noindent To discriminate extrinsic (only due to the loss $\alpha>0$) and intrinsic contribution to irreversibility 
(due to GV) we introduce the fraction of the intrinsic irreversibility  
\begin{equation}
I_{int}(\alpha,P)=1-\frac{R(\alpha,P)}{R(\alpha,0)}\text{.}
\end{equation}
$I_{int}(\alpha,P)=0$ if the irreversibility is only due to linear losses;
if $I_{int}(\alpha,P)>0$ nonlinearity enhances irreversibility and $I_{int}$ quantifies this contribution.

\begin{figure}
  \includegraphics[width=0.45\textwidth]{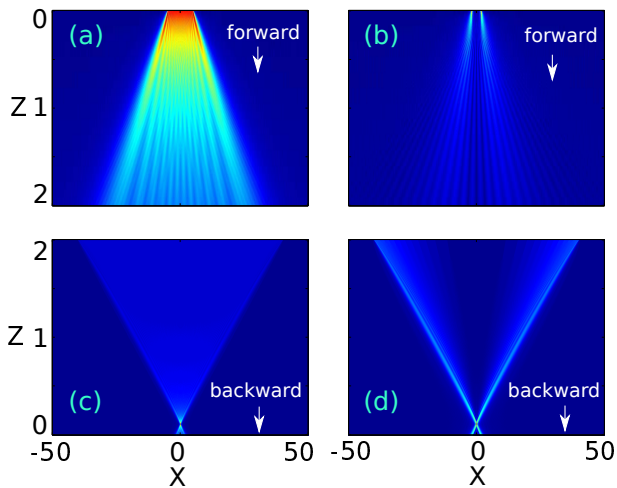}
\caption{
(a) forward propagation of the $\mathfrak{f}_n^-$ truncated GV 
($n=0$, $x_G=40$, $P=4$, $\alpha=0$, $\sigma=10$);
(b) as in (a) for $n=2$;
(c) backward propagation of $\mathfrak{f}_n^+$ truncated GC ($n=0$, $x_G=40$, $P=4$, $\alpha=0$, $\sigma=10$);
(d) as  in (c) for $n=2$.
\label{figureirreversible2}}
\end{figure}

Figure \ref{figureirreversible1}f shows that $I_{int}(\alpha,P)$  grows with $P$. 
To provide evidence of the role of GVs, we consider the evolution according to Eq.(\ref{paraxialnorm}) of the truncated (finite power realization) $\mathfrak{f}_n^{\pm}$ states, i.e., $\psi(x,0)=\mathfrak{f}_n^{\pm}(x)$ for $|x|<x_G$ and zero elsewhere. 
In Fig.\ref{figureirreversible2}a,b we show the dynamics of truncated decaying states $n=0,2$, strongly resembling that observed for a Gaussian initial condition at high power, Fig.~ \ref{coefficientgamow}b, mediating the DSW generation.

\noindent In Fig.\ref{figureirreversible2}c,d we show the backward evolution of the exponentially amplified truncated states $\mathfrak{f}_n^{+}$, which are the time-reversed $\mathfrak{f}_n^{-}$. Albeit no amplification is present, but also losses are included, these undergo a strong field enhancement in propagation and collapse-like dynamics.
This is due to the fact that the power conservation, for negligible loss, forces the exponential shrinking of the wavepacket spatial-width when the peak suffers the exponential increase.
The collapse of the time-reversed GV state is ultimately limited by the nonlocality.
The dynamics of time-reversed GV reproduces that during reversibility breaking in Fig.~\ref{figureirreversible1}d; i.e., the forward evolution generates the $\mathfrak{f}_n^-$ states that, upon time-reversal, becomes the exponentially amplified and collapsing $\mathfrak{f}_n^+=(\mathfrak{f}_n^-)^*$.

\noindent To explain in simple terms the interplay between the linear loss and the GV, we observe that after time reversal the exponentially decaying dynamics of GV, with exponent 
$-\Gamma_n(P)-\alpha$, becomes an exponential amplification of states growing with $\Gamma_n(P')-\alpha$; 
on the contrary, the background component $\phi_b$ is always attenuated.
$P'$ is in general different from $P$ because of linear loss.
The presence of GV radically alters the degree of reversibility because of their exponential dynamics: in a forward/backward round of evolution GV are nearly un-affected by linear loss, while the background component is attenuated.

\noindent The previous arguments allow to derive a simple expression for the degree of irreversibility: at high power the beam is mostly formed by GV; for the $n-$state we have $\psi_F(x)=\mathfrak{f_n}^{-}(x)$.
After forward propagation, letting $L_{eff}=2(1-e^{-\alpha L/2})/\alpha$ we have 
\begin{equation}
  \psi(x,L)=\mathfrak{f_n}^{-}(x)\exp[-\alpha \frac{L}{2}-\Gamma_n(P)\frac{L_{eff}}{2}]\text{.}
\end{equation}
Upon time reversal we have 
\begin{equation}
  \psi'(x,0)=\mathfrak{f_n}^{+}(x)\exp[-\alpha \frac{L}{2}-\Gamma_n(P) \frac{L_{eff}}{2}],
  \end{equation} which propagates again in the forward direction, at a power $P'=P\exp(-\alpha L)$:
\begin{equation}
\psi_B(x)=\mathfrak{f_n}^{+}(x)e^{-\alpha L}e^{\left[\Gamma_n(P')-\Gamma_n(P)\right]\frac{L_{eff}}{2}}
\end{equation}
by projecting over $\psi_F$ we have for $n=0$
\begin{equation}
I_{int}=1-\exp\left[\kappa_2 \sqrt{P} \left(e^{-\alpha \frac{L}{2}}-1\right)\frac{L_{eff}}{2}\right]\text{,}
\label{irreq}
\end{equation}
which is found in good agreement with the numerical values in Fig.\ref{figureirreversible1}e,
also taking into account that we are only retaining the $n=0$ term in the GV expansion.
In the limit of small loss $\alpha$, Eq.(\ref{irreq}) reads as 
\begin{equation}
I_{int}=\frac{\sqrt{P}\alpha L^2}{4 \sqrt[4]{\pi} \sigma}=\frac{\sqrt[4]{\pi}}{4}
\sqrt{\frac{ |n_2| I_0}{n_0}}\frac{ L_Z^2}{ L_{nloc}L_{loss}}\text{,}
\label{irreq1}
\end{equation}
being and $I_0=P_{MKS}/\pi W_0^2$ the peak intensity $L_Z=L Z_d$ the propagation length in real world units. 

The use of GV radically simplifies the treatment of the highly nonlinear and nonlocal regimes.
Eq.(\ref{irreq1}) shows that the strength of nonlinearity has a direct effect on irreversibility, 
due to states with intrinsic exponential decay. A small amount of extrinsic loss leads to a breaking of time-reversal that is amplified by nonlinearity.
In these terms GV have a leading role, as demonstrated by this theoretical analysis that heavily relies on the use of GV exponential trend, which is shown to be in quantitative agreement with numerical simulations.


\section{Conclusions} \label{sec:conc} 
We theoretically predicted Gamow vectors in nonlinear waves and discuss their possible first observation in laser beam propagation in thermal liquids. We discussed their main signature, i.e.,
the quantization of their decay rates, and show the strict link between Gamow vectors and nonlinear wave physics.

In particular, we investigated the degree of irreversibility of dispersive shock waves in highly nonlocal media. We show that the exponentially-decaying dynamics of Gamow vectors enhances the irreversibily, and makes it impossible reverting a dispersive shock wave even in the presence
of a neglible and arbitrarily small loss.
The evidence for this role of Gamow vectors is given by the fact that they allow the analytical calculation of the degree of irreversibility in quantitative agreement with numerical solutions.
This is a subtle result of the interplay between the exponential decay of nonlinear origin and the coupling with environment.

This analysis shows that a nonlocal nonlinear propagation may be used to simulate quantum systems and experimentally study the origin of irreversibility.
Our results are relevant for other problems in nonlinear physics such as rogue waves, beam collapse and filamentation, wave-breaking and supercontinuum generation, Bose-Einstein condensation and polaritons, and also provide support to the RHS formulation of Hamiltonian theories. The exponential dynamics of Gamow vectors may also potentially employed in applications as lasers and nano-focusing of light.

We acknowledge support from the CINECA under the ISCRA initiative, the Progetto di Ateneo Sapienza Award: PhotoAnderson, and the ERC project VANGUARD, grant number 664782.

%

\end{document}